\documentclass[fleqn,usenatbib]{mnras}

\usepackage[english]{babel}
\usepackage[usenames]{color}

\usepackage{newtxtext,newtxmath}
\usepackage[T1]{fontenc}
\usepackage{graphicx}	
\usepackage{amsmath}	

\title[Dynamics of magnetized accretion disks]{Dynamics of magnetized accretion disks of young stars}

\author[S. A. Khaibrakhmanov and A. E. Dudorov]{
Sergey A. Khaibrakhmanov,$^{1,2}$\thanks{E-mail: khaibrakhmanov@csu.ru}
A. E. Dudorov$^{2,1}$\thanks{deceased}
\\
$^{1}$Ural Federal University, 51 Lenin str., Ekaterinburg 620000, Russia\\
$^{2}$Chelyabinsk State University, 129 Br. Kashirinykh str., Chelyabinsk 454001, Russia
}

\date{Accepted 23.07.2022}
\pubyear{2022}

\begin{document}

\label{firstpage}
\pagerange{\pageref{firstpage}--\pageref{lastpage}}
\maketitle

\begin{abstract}
We investigate the dynamics of the accretion disks of young stars with fossil large-scale magnetic field. The author's magnetohydrodynamic (MHD) model of the accretion disks is generalized to take into account the dynamical influence of the magnetic field on gas rotation speed and the vertical structure of the disks.

With the help of the developed dynamical MHD model, the structure of an accretion disk of a solar mass T Tauri star is simulated for different values of the accretion rate $\dot{M}$ and sizes of dust grains $a_{\rm d}$. The simulations of the radial structure of the disk show that the magnetic field in the disk is kinematic, and the electromagnetic force does not affect the rotation speed of the gas for typical values $\dot{M}=10^{-8}\,M_\odot\,\mbox{yr}^{-1}$ and $a_{\rm d}=0.1\,\mu$m. In the case of large dust grains, $a_{\rm d}\geq 1$~mm, the magnetic field is frozen into the gas and a dynamically strong magnetic field is generated at radial distances from the star $r\gtrsim 30$~au, the tensions of which slow down the rotation speed by $\lesssim 1.5$~\% of the Keplerian velocity. This effect is comparable to the contribution of the radial gradient of gas pressure and can lead to the increase in the radial drift velocity of dust grains in the accretion disks. In the case of high accretion rate, $\dot{M}\geq 10^{-7}\,M_\odot\,\mbox{yr}^{-1}$, the magnetic field is also dynamically strong in the inner region of the disk, $r<0.2$~au.

The simulations of the vertical structure of the disk show that, depending on the conditions on the surface of the disk, the vertical gradient of magnetic pressure can lead to both decrease and increase in the characteristic thickness of the disk as compared to the hydrostatic one by 5--20~\%. The change in the thickness of the disk occurs outside the region of low ionization fraction and effective magnetic diffusion (`dead' zone), which  extends from $r=0.3$ to $20$~au at typical parameters.
\end{abstract}

\begin{keywords}
accretion, accretion disks -- Magnetohydrodynamics (MHD) -- Protoplanetary disks -- ISM: magnetic fields
\end{keywords}

\section{Introduction}
\label{sect:intro}

Accretion disks of young stars are geometrically thin rotating gas-dust disks with radii of $\sim 100-1000$~au and masses $\sim 0.001-0.1\,M_\odot$~\citep[see][]{williams11}. During the evolution of accretion disks (ADs), the rate of mass accretion onto the young star decreases from $10^{-6}$ to $10^{-9}\,M_{\odot}\,\mbox{yr}^{-1}$. The accretion disks of young stars can evolve into protoplanetary disks (PPDs), i.~e. analogues of the protosolar nebula, in which conditions are favourable for the planet formation. In recent years, this assumption has received direct observational confirmation~\citep{boccaletti20, nayakshin20}. The processes associated with the formation and dynamics of protoplanets also manifest themselves as rings, spirals, vortices and other substructures in the maps of the emission of the ADs and PPDs in the infrared and (sub)millimetre ranges of the spectrum~\citep{alma15, andrews18}. The development of the techniques for conducting and analysing observations makes it possible to study the structure of the ADs and PPDs, in particular, to restore two-dimensional temperature distributions and determine the velocity components in the disks~\citep[see, for example][]{pinte18, teague21}. The development of theoretical models of the ADs and PPDs is of interest to interpret the existing and future observations of these objects, as well as to explain the origin and properties of extrasolar planetary systems.

Modern observational data allow us to conclude that the ADs and the PPDs of young stars have large-scale magnetic field. The first indications of this were obtained from polarimetric studies of the emission from the vicinity of T Tauri stars~\citep{tamura89}. Polarization mapping of the ADs of T~Tauri and  Herbig Ae/Be stars with the spatial resolution of about 50~au showed that the large-scale magnetic field in the disks can have complex geometry~\citep{li18}. The presence of the magnetic field is indirectly confirmed by the detection of outflows and jets from young stellar objects~\citep[see review by][]{frank14}. Reliable measurements of the magnetic field strength in the ADs and PPDs of young stars are still difficult due to the limitations on the sensitivity and spatial resolution of instruments. \cite{donati05} reported the registration of the Zeeman broadening of the emission lines in the disk of the FU Ori star and claimed that the magnetic field with the intensity of $B\sim 1$~kG may be present in the disk. A promising tool is the measurement of the Zeeman splitting of the CN molecular lines in the sub-mm range~\citep[see][]{kh21fian}, with which it is now possible to get only upper estimates for the magnetic field strength in the surface layers of the disks~\citep{vlemmings19, harrison21}. Finally, measurements of the remnant magnetization of meteorites give indirect estimates of the magnetic field strength of the protosolar nebula $\sim 0.1-1$~G at the distance of $1-3$~au from the protosun~\citep{levy78, fu14}.

The presence of the large-scale magnetic field in the ADs and PPDs is naturally explained within the framework of the theory of a fossil magnetic field. The theory is based on the analysis of observational data and numerical simulations of the star formation as a result of the gravitational collapse of magnetic rotating protostellar clouds (PSCs). According to this theory, the magnetic flux of the PSCs is partially conserved during the collapse of the PSC, and stars with accretion disks born with the fossil large-scale magnetic field ~\citep{dudorov95, dkh15}. Under the condition of developed cyclonic convection in a differentially rotating disk, the generation of the magnetic field due to the dynamo mechanism is also possible~\citep{moss16, sokoloff21}.

The first works on the modelling the evolution of the ADs with the large-scale magnetic field considered the problem of the efficiency of advection of the poloidal magnetic field in the turbulent disks~\citep{lubow94, agapitou96}. \cite{rr96} pointed out that the toroidal magnetic field is also generated in rotating disks. \cite{shu07} derived an analytical solution for the distribution of the poloidal magnetic field in a disk for the prescribed diffusivity and the degree of deviation of the gas rotation velocity from the Keplerian one. \cite{lovelace09} simulated the vertical structure of stationary accretion disks under the assumption that the surface layers of the accretion disks are non-turbulent. In these paper, the diffusivity was prescribed using a dimensionless Prandtl magnetic number. The approach of~\cite{lubow94} was developed in the work of~\cite{guilet14}, in which the influence of the turbulent diffusion on the evolution of the large-scale poloidal magnetic field was postulated and studied. The authors made a careful averaging of the transport coefficients over the height of the disk and took into account the self-consistent influence of the magnetic field on these coefficients, which made it possible to describe a more effective amplification of the magnetic field in the disks. \cite{okuzumi14} obtained a stationary distribution of the vertical magnetic field in the disk, taking into account only Ohmic dissipation. \cite{lizano16} simulated the vertical structure of the AD of a young star with a prescribed frozen-in poloidal magnetic field. They examined the effect of viscous and Ohmic heating, as well as the heating by stellar radiation on the temperature of the disk.

Most of the numerical MHD simulations of the evolution of the ADs taking into account Ohmic dissipation and magnetic ambipolar diffusion~\citep{simon13}, as well as taking into account the Hall effect~\citep{lesur14}, were carried out only in a local approximation using the prescribed strength and/or geometry of the magnetic field. \cite{gressel15} performed the first global simulations taking into account magnetic ambipolar diffusion and ionization  by X-rays and ultraviolet radiation of the star only. The simulations investigated the dynamics of the disk in a small region from 1 to 5 au from the star. In a number of recent works, large-scale MHD simulations of the evolution of the ADs and PPDs have been carried out in order to study the  angular momentum transport through disk wind and to explain the formation of ring gas structures due to MHD processes~\citep{bethune17, suriano18, gressel20, riols20}. These works used a priori fixed magnetic diffusivity and/or  magnetic field strength under the assumption of the constant plasma beta in the entire disk.

Within the framework of the theory of the fossil magnetic field, \cite{dkh13a, dkh13b, dkh14, kh17} developed a kinematic MHD model of the accretion disks of young stars. The model is based on the approximations of the standard model of \cite{ss73}. In addition to the equations of the basic model, we solve the induction equation taking into account Ohmic dissipation and magnetic ambipolar diffusion, magnetic buoyancy and the Hall effect, as well as the equations of thermal and shock ionization taking into account the main ionization and recombination effects. Simulations using the model have shown in particular that the magnetic field can be dynamically strong in some regions of the disk, $\beta\lesssim 1$, where $\beta$ is the plasma beta (the ratio of gas pressure to magnetic pressure). In this regard, it is necessary to develop a dynamical model of the ADs, which would take into account the influence of the magnetic field on the structure and evolution of  disk. Earlier, the Dudorov and Khaibrakhmanov's model was extended to account for the effect of dissipative MHD effects on the thermal structure of the disk ~\citep{kh18mhd}, as well as the magnetic pressure gradient on the vertical structure of the disk~\citep{cpmj21}. In this paper, we carry out further development of the model. For the first time, the effect of large-scale magnetic field tensions on the centrifugal equilibrium of the disk is considered in a self-consistent manner and the degree of deviation from the Keplerian rotation is determined.

The paper is organized as follows. Section~\ref{sect:problem} describes the problem statement. The equations of the model describing the radial and vertical structures of the disk are written in sections~\ref{sect:eq_radial} and \ref{sect:eq_vertical}, respectively. Section~\ref{sect:params} provides the main parameters of the model and methods for solving the equations. The results of the simulations of the radial structure of the disk and the influence of the magnetic field on the rotation speed of the disk are analysed in section~\ref{sec:radial}. Section~\ref{sec:vertical} describes the results of the simulations of the vertical structure of the disk and the effect of the magnetic field on the thickness of the disk. Section~\ref{sect:end} discusses the results of the work and draws conclusions.

\section{Problem statement}
\label{sect:problem}

Consider a stationary geometrically thin and optically thick accretion disk, whose mass is small compared to the mass of a star. We use cylindrical coordinates $(r,\,\varphi,\,z)$. The angular momentum is assumed to be transferred in the radial direction $r$ by turbulence. The disk is in a state of centrifugal equilibrium in the $r$-direction and magnetostatic equilibrium in the vertical direction $z$. The energy released in the disk due to turbulent friction is carried away by the radiation in the vertical direction.

The matter in the ADs of young stars is weakly ionized gas-dust plasma. To study the dynamics of the AD with large-scale magnetic field, we use the system of equations of dissipative magnetic gas dynamics,

\begin{eqnarray}
  \frac{\partial\rho}{\partial t} + {\bf \nabla}\cdot\left(\rho\textbf{v}\right) &=& 0, \label{Eq:Cont}\\
  \rho\left[\frac{\partial {\bf v}}{\partial t} + \left({\bf v}\cdot \nabla\right){\bf v}\right] &=& -\nabla \left(p + \frac{B^2}{8\pi}\right) + \rho{\bf g} + \frac{1}{4\pi}\left({\bf B}\cdot \nabla\right){\bf B} \nonumber \\
& & + \, \mathrm{div}\hat{\sigma}',\label{Eq:Motion}\\
  \rho T\left[\frac{\partial s}{\partial t} + \left({\bf v}\cdot {\bf \nabla}\right)s\right] &=& \sigma_{ik}^{\prime}\frac{\partial v_i}{\partial x_k} + {\bf \nabla}\cdot \mathcal{\textbf{F}},\label{Eq:TotalEnergy}\\
  \frac{\partial\textbf{B}}{\partial t} &=& \nabla\times \left(\textbf{v} \times \textbf{B}\right) + \eta \nabla^2\textbf{B}, \label{Eq:Induction}
\end{eqnarray}
where $\textbf{g}$ is the gravitational acceleration, $s$ is the entropy per unit volume, $\hat{\sigma}'$ is the viscous stress tensor ($\sigma_{ik}^{\prime}$ in the index notation, the indices $i$ and $k$ number the spatial coordinates), $\mathcal{\textbf{F}}$ is the the radiative energy flux density, $\eta$ is the the magnetic diffusivity, including Ohmic dissipation and magnetic ambipolar diffusion. Standard physical notations are used for the remaining values. To close the system of equations (\ref{Eq:Cont}--\ref{Eq:Induction}), we use the equation of state of the ideal gas with mean molecular weight of $\mu=2.3$.

\section{Equations of the MHD model of accretion disks}

In the approximation of the geometrically thin disk, radial gradients can be neglected in comparison with the vertical one, and the latter can be replaced by the finite differences in the initial equations. In the considered case of a low-mass disk, the gravitational acceleration is determined by the star,
\begin{equation}
\textbf{g} = -\frac{GM}{R^3}\textbf{R},
\end{equation}
where $\textbf{R} = (r,\, 0,\, z)$ is the radius vector. In the geometrically thin disk, the $r$-component of the gravitational force dominates, so that the following
relations are fulfilled for the components of the gas velocity vector: $|v_z|\ll |v_r| \ll 
|v_\varphi|$. These approximations allow us to separate the variables in the MHD equations and solve the radial structure equations (accretion equations) and the vertical structure equations (magnetostatic equilibrium equation) separately, assuming that the disk is in centrifugal equilibrium.

The applicability of the stationary approximation is validated by the fact that the main dynamic time scale in the considered case is the Keplerian period, $t_{\rm k} = 1\,\mbox{yr}\left(r/\mbox{1 au}\right)^{3/2}$, which is  small compared to the lifetime of the disk ~$t_{\rm disk}\sim 1-10$~million yrs.

\subsection{Equations of the radial structure}
\label{sect:eq_radial}

The system of equations describing the radial structure of the disk in the kinematic approximation has been derived by~\cite{dkh13a, dkh13b, dkh14}. Let us consider how the equations of the system change if one take into account the dynamical terms, namely: the influence of magnetic tensions on centrifugal equilibrium and dissipative MHD effects on the thermal balance in the disk.

The influence of the magnetic field on the rotation speed of the disk is described by the radial component of the motion equation~(\ref{Eq:Motion}), 
\begin{eqnarray}
\rho\Omega^2r - \rho \frac{GM}{r^2}\left(1 + \frac{z^2}{r^2}\right)^{-3/2} + \frac{\partial p}{\partial r} & & \nonumber\\
 + \frac{B_z}{4\pi}\frac{\partial B_r}{\partial z} - \frac{\partial}{\partial r}\left(\frac{B_z^2+B_\varphi^2}{8\pi}\right) - \frac{B_\varphi^2}{4\pi r} &=& 0,\label{eq:cfb}
\end{eqnarray}
where $\Omega = v_\varphi/r$ is the the angular velocity of the gas. The gas pressure gradient, $\partial p/\partial r$, in a geometrically thin disk can also be neglected. Note that this term plays an important role in the problems of the interaction of gas and dust grains in the disks, because the small deviation from the centrifugal balance caused by it leads to the drift of dust grains relative to the gas~\citep{w77}. The fourth term in the equation (\ref{eq:cfb}) describes magnetic tensions, and the last two correspond to the magnetic pressure gradient. In the geometrically thin disk, the last two terms can be neglected, therefore
\begin{equation}
v_\varphi = \sqrt{\frac{GM}{r}\left(1 + \frac{z^2}{r^2}\right)^{-3/2} - \frac{rB_z}{4\pi\rho}\frac{\partial B_r}{\partial z}}.
\end{equation}
This equation shows that the main contribution of the electromagnetic force to the centrifugal balance is determined by the $rz$-component of the Maxwell stress tensor. In the model approximations, this term can be estimated as $(rB_z/4\pi\rho)(B_r/H)$.

The effect of the dissipative MHD effects~--- Ohmic dissipation and magnetic ambipolar diffusion~--- on the thermal structure of the disk was included in the model and investigated by~\cite{kh18mhd}.

The final system of equations of the radial disk structure, taking into account the considered dynamical effects, can be written as follows:
\begin{eqnarray}
\dot{M} &=& -2\pi rv_r\Sigma,\label{eq:Mdot}\\
\dot{M}\Omega_{\rm k}f &=& 2\pi\alpha c_{\rm T}^2 \Sigma,\label{eq:ang_mom}\\
v_\varphi &=& \sqrt{\frac{GM}{r}\left(1 + \frac{z^2}{r^2}\right)^{-3/2} - \frac{rB_rB_z}{4\pi\Sigma}},\label{eq:vphi}\\
H &=& \frac{c_{\rm T}}{\Omega_{\rm k}},\label{eq:H}\\
\sigma_{\rm sb}T_{\rm eff}^4 &=& \frac{3}{8\pi}\dot{M}\Omega_{\rm k}^2f + \Gamma_{\rm MGD},\label{eq:Teff}\\
T^4 &=& \frac{3}{8}\kappa_{\rm R}\Sigma T_{\rm eff}^4,\label{eq:Tc}\\
B_r &=& -\frac{v_rH}{\eta}B_z,\label{eq:Br}\\
B_\varphi &=& -\frac{3}{2}\left(\frac{H}{r}\right)^2\frac{v_\varphi H}{\eta}B_z - \frac{1}{2}\left(\frac{H}{r}\right)\frac{v_\varphi H}{\eta}B_r,\label{eq:Bphi}\\
B_z &=& \left\{ 
\begin{array}{rcl}
B_{z0}\dfrac{\Sigma}{\Sigma_0},\quad {R}_{\rm m} \gg 1\\
\sqrt{4\pi x\rho^2 r|v_r|}, \quad {R}_{\rm m} < 1\\
\end{array}
\right.\label{eq:Bz}
\end{eqnarray}
where $\dot{M}$ is the accretion rate, $\Sigma = 2\langle\rho\rangle H$ is the  gas surface density, $\langle\rho\rangle$ is the $z$-averaged  gas density, $H$ is the disk  scale height, $\Omega_{\rm k}=\sqrt{GM/r^3}$ is the Keplerian angular velocity, $f =1 - \left(r_0/r\right)^{1/2}$, $r_0$ is the radial coordinate of the inner boundary of the disk, $c_{\rm T}=\sqrt{R_{\rm g}T/\mu}$ is the isothermal sound speed, $T$ is the temperature in the equatorial plane of the disk, $\sigma_{\rm sb}$ is the Stefan-Boltzmann constant, $T_{\rm eff}$ is the  disk's effective temperature, $\Gamma_{\rm MGD}$ is the heating rate due to dissipative MHD effects per unit area of the disk, $\kappa_{\rm R}$ is the Rosseland mean opacity, $B_{z0}=B_z(r_{\rm out})$, $\Sigma_0=\Sigma(r_{\rm out})$, $r_{\rm out}$ is the outer radius of the disk, $R_{\rm m}$ is the magnetic Reynolds number, $x$ is the ionization fraction.

Let us briefly discuss the physical meaning of the model equations. Equation (\ref{eq:Mdot}) follows from the continuity equation. The equation (\ref{eq:ang_mom}) is the angular momentum transfer equation. 
The equation (\ref{eq:Teff}) reflects the balance between the surface heating rates due to turbulent friction (the first term on the right-hand side) and due to dissipative MHD effects (the second term) and the radiation flux. The surface heating rate due to Ohmic dissipation and magnetic ambipolar diffusion is defined as~\citep[see][]{kh18mhd}:
\begin{equation}
\Gamma_{\rm MGD} = \frac{\nu_{\rm m}}{4\pi}\frac{B_r^2 + B_\varphi^2}{H} + \frac{\left(B_rB_z\right)^2 + \left(B_\varphi B_z\right)^2 + \left(B_r^2 + B_\varphi^2\right)^2}{32\pi^2R_{\rm in}H},
\end{equation}
where $\nu_{\rm m}$ is the Ohmic diffusivity, $R_{\rm in}$ is the coefficient of friction between ions and neutrals in plasma.

Relation (\ref{eq:Tc}) between the temperature in the equatorial plane of the disk and the effective temperature is obtained from the solution of the radiation transfer equation in the diffusion approximation. Solutions (\ref{eq:Br}, \ref{eq:Bphi}) of the induction equation for the magnetic field components $B_r$ and $B_\varphi$ reflect the balance between the advection of the magnetic field in the corresponding directions and its diffusion in the $z$-direction. The contribution of the Hall effect to the induction equation was considered by~\cite{kh17}, and it is not taken into account  in this paper for clarity. Solutions (\ref{eq:Bz}) for the vertical component of the magnetic field $B_z$ are derived for the cases of frozen-in magnetic field (i.~e. ideal MHD, $R_{\rm m}\gg 1$) and effective magnetic ambipolar diffusion ($R_{\rm m}<1$).

The Ohmic and magnetic ambipolar diffusivities depend on the ionization fraction and are calculated in accordance with the work of~\cite{dkh14}. The degree of ionization is calculated following~\cite{dudsaz87}. Thermal ionization of metals and hydrogen, shock ionization by cosmic rays, X-rays and radioactive elements are considered. Radiative recombinations and recombinations on dust grains are taken into account.

\subsection{Equations of the vertical structure}
\label{sect:eq_vertical}

The equations of the vertical structure of the AD with large-scale magnetic field are derived by~\cite{cpmj21}. It is assumed that the disk is in magnetostatic equilibrium and the main contribution of the electromagnetic force to the forces balance in the $z$-direction is determined by the gradient of the toroidal magnetic field. The magnetostatic equilibrium is established on the time scale of the propagation of MHD waves in the $z$-direction,
\begin{equation}
t_{\rm A}\approx \frac{H}{\sqrt{c_{\rm T}^2 + v_{\rm A}^2}}
\sim \Omega_{\rm k}^{-1}\left(1 + \frac{2}{\beta}\right)^{-1/2},\label{eq:t_A}
\end{equation}
where $v_{\rm A}$ is the the Alfv{\' e}n velocity. Estimate~(\ref{eq:t_A}) shows that $t_{\rm A}\sim t_{\rm k}\ll t_{\rm disk}$ at $\beta\sim1$.

It is assumed that the heat released as a result of the turbulent friction of adjacent gas layers is carried away by radiation in the vertical direction. For an optically thick disk, the radiation flux is written in the diffusion approximation. In the adopted approximations, the magnetostatic equilibrium equations are following:
\begin{eqnarray}
\frac{d p}{d z} &=& -\rho \frac{GM}{r^3}z - \frac{d}{d z}\left(\frac{B_{\varphi}^2}{8\pi}\right),\label{eq:motion_z3}\\
-\frac{16\sigma T^3}{3\kappa_{\rm R}\rho}\frac{d T}{d z} &=&  \mathcal{F}_z,\label{eq:dT_dz1}\\
\frac{d \mathcal{F}_z}{d z} &=& \frac{3}{2}\alpha p\Omega_{\rm k},\label{eq:dQturb_dz}\\
\frac{d^2 B_\varphi}{d z^2} &=& -\frac{3}{2}\frac{v_{\rm k}B_z}{\eta}\frac{z}{r^2}, \label{eq:induction_phi}
\end{eqnarray}
where $\mathcal{F}_z$ is the $z$-component of the radiative flux density vector $\textbf{F}$, $v_{\rm k} = \Omega_{\rm k}r$ is the Keplerian speed.

The first-order ordinary differential equations (\ref{eq:motion_z3}), (\ref{eq:dT_dz1}) and (\ref{eq:dQturb_dz}) and the second-order equation (\ref{eq:induction_phi}) form a closed system of equations with respect to the variables $p$, $T$, $\mathcal{F}_z$ and $B_{\varphi}$ for given coefficients $\kappa_{\rm R}(\rho,\,T)$, $v_{\rm k}(r)$, $\eta(r)$ and $B_z(r)$ at some radius of $r$.  We solve the equations in the domain $z\in [0,\, z_{\rm s}]$, where $z_{\rm s}$~is the coordinate of the photosphere of the disk, which is characterized by the optical thickness $\tau=\int_{z_{\rm s}}^{\infty}\kappa_{\rm R}\rho dz=2/3$. Boundary conditions for pressure, temperature and radiative flux density:
\begin{eqnarray}
p (z_{\rm s}) &=& p_{\rm s},\label{eq:BC_p}\\
T(z_{\rm s}) &=& T_{\rm eff},\label{eq:BC_T}\\
\mathcal{F}_z(0) &=& 0,\label{eq:BC_Q0}\\
\mathcal{F}_z(z_{\rm s}) &=& \sigma T^4_{\rm eff}.\label{eq:BC_Qs}
\end{eqnarray}
The pressure at the photosphere boundary $p_{\rm s}$ can be estimated from an approximate solution of the hydrostatic equilibrium equation near the photosphere~\citep[see][]{ malanchev15, ShakuraBook}
\begin{equation}
p_{\rm s} = \frac{2}{3}\frac{\Omega_{\rm k}}{\kappa_{\rm R}(\rho_{\rm s},\,T_{\rm eff})}z_{\rm s}.\label{eq:p_s}
\end{equation}
It is assumed that $p(\tau=0) \ll p_{\rm s}$ when obtaining this estimate.

Due to the equatorial symmetry of the fossil large-scale magnetic field, $B_\varphi(z=0)=0$.

Both Dirichlet (I) and Neumann (II) boundary conditions can be considered at the disk surface for the $\varphi$-component of the magnetic field.

In the case of type I boundary conditions at the disk's surface, the magnetic field strength is fixed
\begin{equation}
 B_\varphi(z_{\rm s}) = B_{\rm ext}, \label{eq:BC_I}
\end{equation}
where the value of $B_{\rm ext}$ is determined by the model of the medium above the disk. For example, if the outflow forms over the disk due to the action of the toroidal magnetic field gradient~\citep{kudoh97}, then $B_{\rm ext}\sim B_z$ and $\beta\sim 1$. In the case of the magneto-centrifugal outflow of~\cite{bp82}, the magnetic field is quasi-radial and $B_\varphi < B_r\sim B_z$ at the surface of the disk.

In the case of type II boundary conditions, it is assumed that the medium above the disk is vacuum,
\begin{equation}
\partial B_\varphi/\partial z (z_{\rm s}) = 0.\label{eq:BC_II}
\end{equation}

\section{Simulation results}
\label{sect:results}

\subsection{Parameters of the model and solution methods}
\label{sect:params}
System of equations (\ref{eq:Mdot}--\ref{eq:Bz}, \ref{eq:motion_z3}--\ref{eq:induction_phi}) together with the equations for calculating the ionization fraction form closed system of equations describing the radial and vertical structure of the disk. These equations are solved in the following sequence.

At the first step, system of nonlinear algebraic equations (\ref{eq:Mdot}--\ref{eq:Bz}), which describes the radial structure of the disk, is solved by the iteration method together with the bisection method. The solution of the corresponding equations in the kinematic approximation~\citep{dkh14} is chosen as the initial approximation. It is assumed that the  magnetic field strength is limited by the maximum value corresponding to the equipartition of magnetic energy and thermal energy of the gas, $B_{\rm eq}=\sqrt{8\pi p}$. The equations are solved in the region from the inner boundary of the disk, $r_{\rm in}$, to the outer boundary, $r_{\rm out}$, on the logarithmic grid containing 200 nodes. The inner boundary of the disk is determined by the radius of the magnetosphere of the star, the outer boundary is the contact discontinuety with the interstellar medium. The magnetic field strength at the outer boundary, $B_{z0}$, is matched to the magnetic field strength of the interstellar medium~\citep[see][]{dkh14}.  The main parameters for the radial structure equations are the accretion rate and turbulence parameter, as well as the characteristics of the star. The parameters of a solar mass T Tauri star are chosen as standard in this work: $\alpha=0.01$, $\dot{M}=10^{-8}\,M_{\odot}\,\mbox{yr}^{-1}$, radius of the star $R_{\star}=2\,R_{\odot}$, luminosity of the star $L_{\star}=1\,L_{\odot}$, magnetic field strength on the surface of the star $B_{\star}=2$~kG. The degree of ionization and magnetic diffusivities are calculated for the following standard parameters: the radius of a dust grain $a_{\rm d}=0.1\,\mu$m, the ionization rate by unattenuated cosmic rays $\xi_0=3\times 10^{-17}\,\mbox{s}^{-1}$, attenuation depth for the cosmic rays $\Sigma_{\rm CR}=100\,\mbox{g cm}^{-2}$, ionization rate due to radionuclides decay $\xi_{\rm RE}=7.6\times 10^{-19}\,\mbox{s}^{-1}$, X-ray luminosity of the star $L_{\rm XR}=10^{30}\,\mbox{erg}\,\mbox{s}^{-1}$, X-ray photon energy $kT_{\rm XR}=0.5$~keV. The opacity coefficient $\kappa_{\rm R}$ is calculated using the interpolation of the tables of~\cite{semenov03} and OPAL~\citep{opal}.

At the second step, ordinary differential equations (\ref{eq:motion_z3}-\ref{eq:induction_phi}) describing the vertical structure of the disk are solved at each given $r$ by the Runge-Kutta method of the 4th order of accuracy with automatic step control. The coefficients of the equations are adapted from the solution of the equations of the radial structure. The coordinate of the photosphere is found by the shooting method. It is assumed that $B_{\rm ext}=-0.1\,B_z$ in the case of type I boundary conditions for the magnetic field at the disk's surface.

Iterative procedures at each step of simulation are carried out for the required relative accuracy of $10^{-4}$. The algorithm for solving the model equations is implemented in the `Belmondo' code written in the C++ programming language.

\subsection{Radial structure of the disk}
\label{sec:radial}
This section presents the results of the simulations of the radial structure of the disk for various parameters and analyses the influence of the magnetic field on the deviation of gas rotation law from the Keplerian one.

\subsubsection{Kinematics}
We first consider the case of a solar mass T Tauri star with standard parameters.

\begin{figure*}
\includegraphics[]{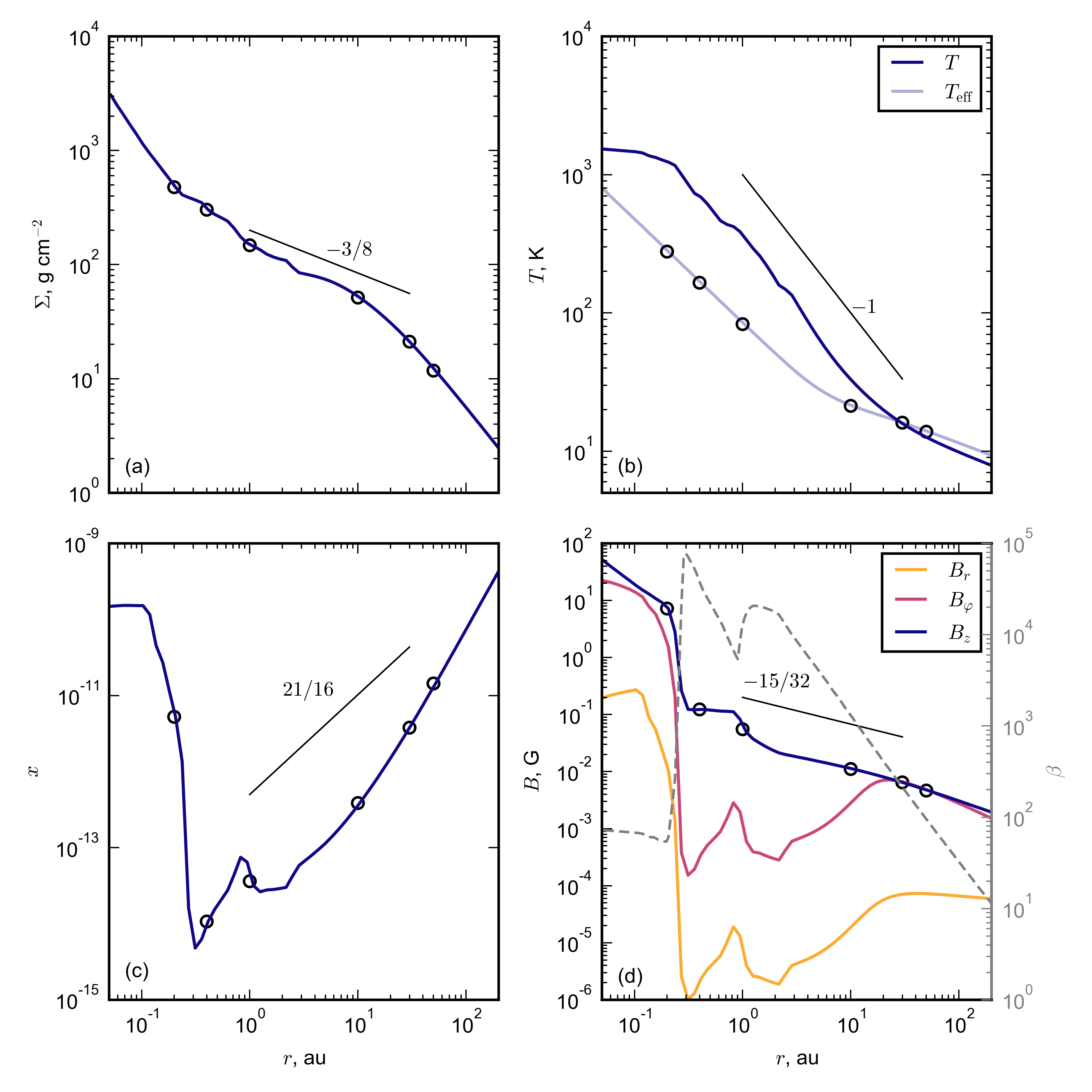}
\caption{The radial structure of the accretion disk of a solar mass T Tauri star with standard parameters. Panel (a): surface density profile. Panel (b): profiles of effective temperature $T_{\rm eff}$ and temperature in the equatorial plane of the disk. Panel (c): ionization degree profile. Panel (d): profiles of the components of the magnetic induction vector (left $y$-axis, solid lines) and the plasma beta (right $y$-axis, dashed line). The circles on the lines mark the points for which the vertical structure is modelled (see figures ~\ref{fig:z} and ~\ref{fig:zphot} below). Lines with numbers indicate characteristic profiles from the analytical solution of the model equations~\citep{dkh14}.}
\label{fig:rad_fiduc}
\end{figure*}

Let us briefly discuss the disk's structure in this case~\citep[for a detailed analysis, see][]{dkh14, kh17}. Figure ~\ref{fig:rad_fiduc} shows the radial profiles of the surface density, temperature, ionization fraction and components of the magnetic induction vector, as well as the profile of the plasma  $\beta$.

Figures~\ref{fig:rad_fiduc}(a) and (b) show that the density and temperature decrease with distance in the disk. In the region $1<r<30$~au, the characteristic slopes of the profiles $\Sigma(r)$ and $T(r)$ are $-3/8$ and $-1$, respectively, which is consistent with typical observational values~\citep[see, for example,][]{williams11}. The disk is optically thick with respect to its own radiation and $T>T_{\rm eff}$ in the region of $r\lesssim 50$~au.

According to Figure~\ref{fig:rad_fiduc}(c), the profile of the ionization fraction is non-monotonic and has a minimum of $x_{\rm min}\approx 5\times 10^{-15}$ near $r_{\rm min}\approx 0.3$~au. In the region of $r<r_{\rm min}$, the ionization fraction increases sharply when approaching the star due to the thermal ionization of metals. In the region $r>r_{\rm min}$, the ionization fraction increases with the distance from the star due to the decrease in the density and more effective ionization by cosmic rays and stellar X-rays.

Figure~\ref{fig:rad_fiduc}(d) shows that three regions with different magnetic field geometries can be distinguished in the disk. In the region $r<r_{\rm min}$, the magnetic field is quasi-azimuthal, $B_r <B_\varphi\sim B_z$. The magnetic field is frozen into the gas, so there is intensive generation of the toroidal magnetic field in this region of high ionization fraction. In the region $r_{\rm min} < r\lesssim 20$~au, the magnetic field is quasi-uniform, $B_z\gg (B_r,\, B_\varphi)$. This region is a `dead' zone in which the ionization fraction is small, $x\lesssim 10^{-12}$, and Ohmic dissipation prevents the magnetic field amplification. Outside the `dead' zone, $r>20$~au, the magnetic field is quasi-azimuthal. Effective magnetic ambipolar diffusion suppresses the amplification of the radial component of the magnetic field in this region. The magnetic field can be quasi-radial, $B_r\sim B_z$, in this region under the condition of increased ionization rates~\citep{dkh14} or due to the Hall effect~\citep{kh17}.

The radial profile of the plasma $\beta$ shows that the magnetic field in the entire disk is kinematic, $\beta\gg 1$, i.~e. the electromagnetic force does not affect the structure of the disk. Note that the value of the plasma beta varies throughout the disk, from $\beta\sim 100$ near the inner boundary of the disk to $\beta\sim (10^3-10^5)$ in the `dead' zone and $\beta\rightarrow 10$ at the disk's periphery.

\subsubsection{Dynamics}
The most intense generation of the magnetic field occurs in the regions with high ionization fraction, i.~e. outside the `dead' zone. To study the conditions for the generation the dynamically strong magnetic field, consider the limiting case when there is no `dead' zone in the disk. As it has been shown by~\cite{dkh14}, this case is realized in the AD with high accretion rate, $\dot{M}\geq 10^{-7}\,M_\odot\,\mbox{yr}^{-1}$, and/or in the cases when there are no dust grains in the disk or the dust grains have large sizes.

\begin{figure*}
\includegraphics[]{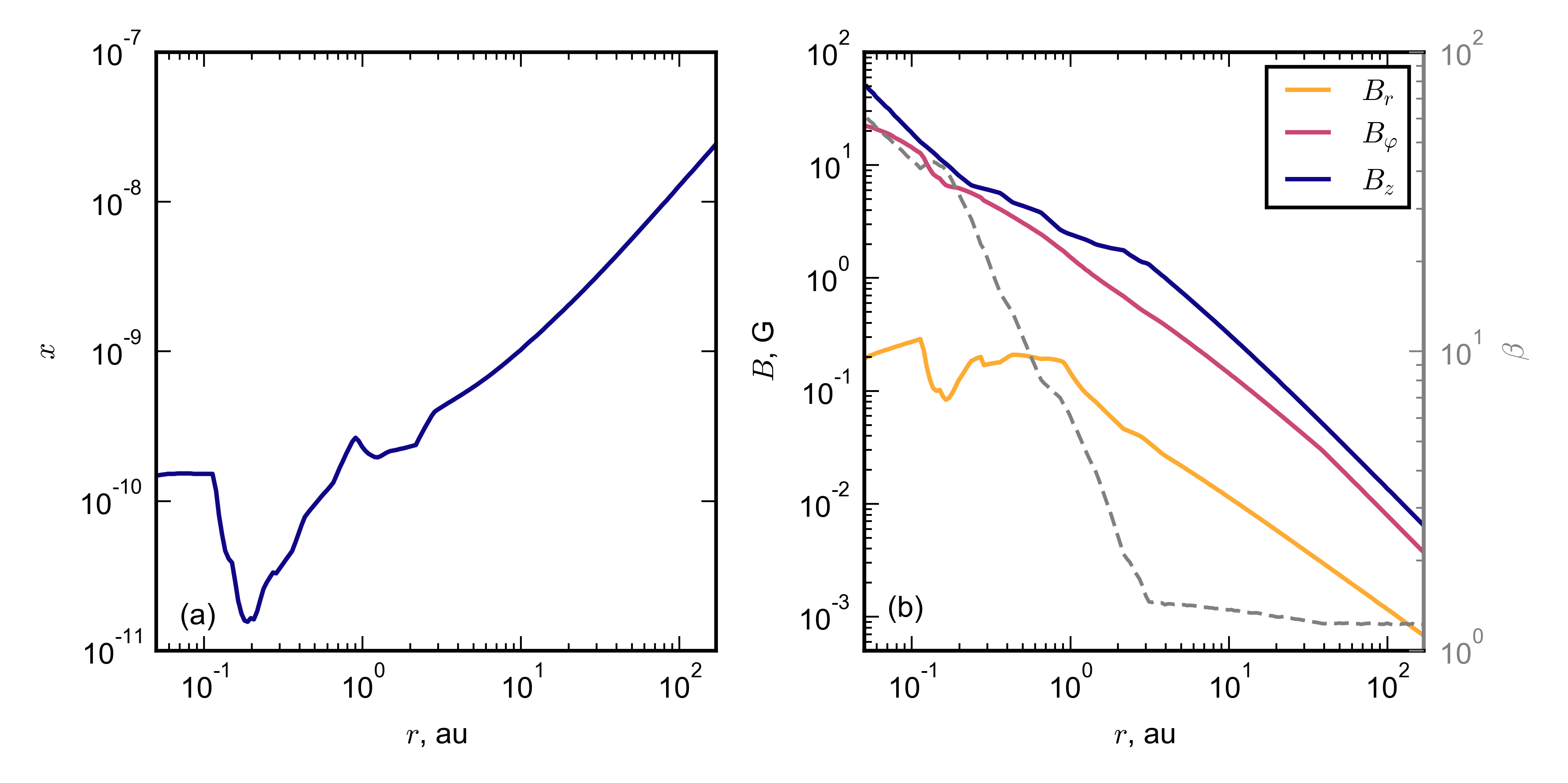}
\caption{Same as in Fig.~\ref{fig:rad_fiduc}(c) and (d), but for the case of dust grains of size $a_{\rm d}=1$~mm.}
\label{fig:rad_mgd}
\end{figure*}

Figure~\ref{fig:rad_mgd} shows the radial profiles of the ionization fraction and the magnetic field components in the disk with the same parameters as in Figure~\ref{fig:rad_fiduc}, but for the case of dust grains with the size of $1$~mm.

According to figure~\ref{fig:rad_mgd}(a), the radial profile of the ionization fraction is non-monotonic, as in the case of $a_{\rm d}=0.1\,\mu$m. 
The difference is that the minimum value of $x$ is $\sim 10^{-11}$, which is five orders of magnitude higher than in the case discussed in the previous section. This is explained by the fact that the main type of recombinations are recombinations on dust grains in the considered range of densities, and $x\propto a_{\rm d}$ in this case ~\cite[see][]{dkh14}.  Thus, there is no `dead' zone in the case of $a_{\rm d}\gtrsim 1$~mm.

Due to the higher ionization fraction than in the case of small dust grain, intense magnetic field generation occurs throughout the disk. The component $B_z$ in this case is proportional to the surface density of the disk $\Sigma$. As Figure~\ref{fig:rad_mgd}(b) shows, the intensity of $B_z$ decreases from about $500$~G at the inner edge of the disk to $25$~G at $r=1$~au and $0.06$~G at the outer boundary of the disk, $r_{\rm out}=165$~au. The figure shows that the magnetic field is quasi-azimuthal, $B_\varphi\sim B_z$, in the entire disk. The intensity of the radial component of the magnetic field is about an order of magnitude less than the intensity $B_z$ in the region $r>1$~au and by two orders of magnitude in the region of the minimum ionization fraction, $r<1$~au.

The radial profile of the plasma beta shows that the magnetic field is dynamically strong at the periphery of the disk: $\beta\sim 1$ at $r>30$~au. The plasma beta increases as it approaches the star in the inner region of the disk, $r<10$~au, and it reaches $\beta\sim 50$ at the inner boundary of the disk.

\subsubsection{Deviation from the Keplerian rotation}
The results presented in Figure~\ref{fig:rad_mgd} showed that the dynamically strong magnetic field is generated in the disk if the dust grains in the AD are large, $a_{\rm d}\gtrsim 1$~mm. We can expect a noticeable effect of magnetic tensions on the angular velocity of the gas  in this case. This effect is described by the second term under the root sign in the equation (\ref{eq:vphi}). Consider the contribution of magnetic tensions to the deviation of the gas velocity from the Keplerian one, and compare this contribution with the deviation caused by the gas pressure gradient. If both effects are taken into account, then the following expression can be obtained from the centrifugal balance equation~(\ref{eq:cfb}) for the azimuthal velocity of the gas in the disk:
\begin{equation}
v_\varphi = \sqrt{\frac{GM}{r}\left(1 + \frac{z^2}{r^2}\right)^{-3/2} - \frac{r}{\rho}\frac{\partial p}{\partial r}- \frac{rB_rB_z}{2\pi\Sigma}}.
\end{equation}
This formula can be rewritten in the following form
\begin{equation}
v_\varphi = v_{\rm k}\sqrt{1 - \beta_{\rm k}^{\rm GD} - \beta_{\rm k}^{\rm MGD}},\label{eq:v_subkepler}
\end{equation}
where
\begin{equation}
\beta_{\rm k}^{\rm GD} = \left( \frac{r}{\rho}\frac{\partial p}{\partial r}\right)\frac{1}{g_r}, \quad \beta_{\rm k}^{\rm MGD} = \left(\frac{rB_rB_z}{2\pi\Sigma}\right) \frac{1}{g_r} \label{eq:beta_k}
\end{equation}
are the degrees of the deviation of the gas velocity from the Keplerian one due to the action of the gas pressure gradient and magnetic tensions, respectively. It follows from the formula ~(\ref{eq:v_subkepler}) that the difference between the Keplerian velocity and $v_\varphi$ is $\Delta v\approx 1/2\beta_{\rm k}v_{\rm k}$, where $\beta_{\rm k}$~is one or both of the values in ~(\ref{eq:beta_k}). For example, if $\beta_{\rm k}=0.02$, one get $\Delta v \approx 1\,\%\,v_{\rm k}$.

\begin{figure*}
\includegraphics[]{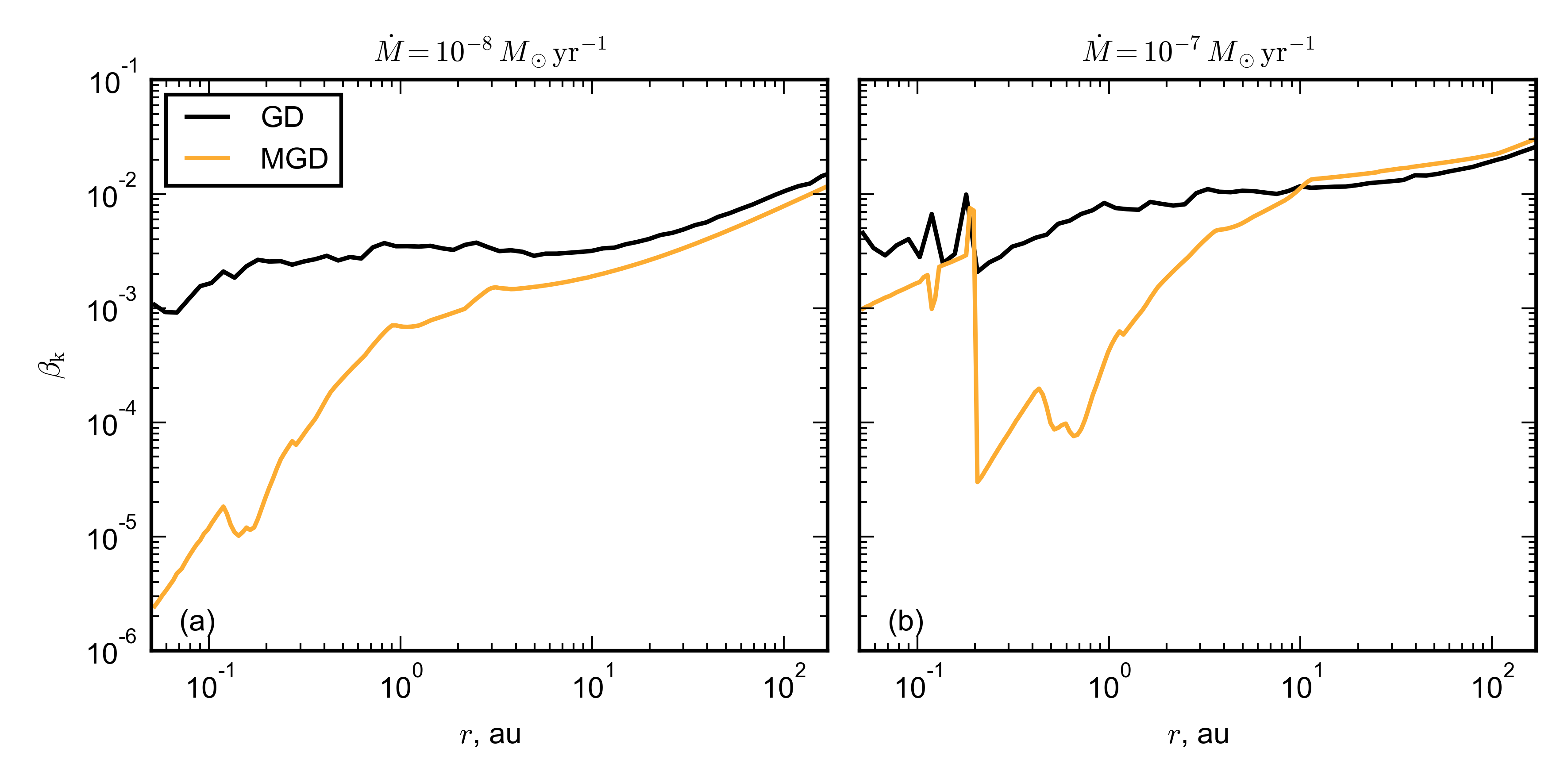}
\caption{The radial profiles of the degrees of the deviation of the gas azimuthal velocity from the Keplerian one~(\ref{eq:beta_k}). The black line (GD) shows the deviation due to the gas pressure gradient, the yellow line (MGD) corresponds to the effect by magnetic tensions. The simulations were carried out for the radius of dust grain $a_{\rm d}=1$~mm. Panel~(a): the accretion rate is $10^{-8}\,M_\odot\,\mbox{yr}^{-1}$ (corresponds to figure~\ref{fig:rad_mgd}), panel~(b): accretion rate~$10^{-7}\,M_\odot\,\mbox{yr}^{-1}$.}
\label{fig:subkepler}
\end{figure*}

Figure~\ref{fig:subkepler}(a) shows the radial profiles of the quantities~(\ref{eq:beta_k}) for the simulation with standard parameters at $a_{\rm d}=1$~mm. Figure~\ref{fig:subkepler}(b) shows similar profiles for the simulation with the same parameters, but for the accretion rate of $10^{-7}\,M_\odot\,\mbox{yr}^{-1}$. The derivative of the gas pressure by $r$ in the expression $\beta_{\rm k}^{\rm GD}$ is calculated from profile $p(r)$ using the numerical quadrature.

Figure ~\ref{fig:subkepler}(a) shows that the `gas-dynamic' (GD) degree of the deviation of the gas angular velocity from the Keplerian one, $\beta_{\rm k}^{\rm GD}$, varies little with distance, and its value lies in the range from $10^{-3}$ near the inner edge of the disk up to $2\times 10^{-2}$ ($\Delta v\approx 1\%\,v_{\rm k}$) at the outer edge of the disk. The characteristic value of $\beta_{\rm k}^{\rm GD}$ in the region of $r\in[1,\,50]$~au is $(3-4)\times 10^{-3}$ ($\Delta v \approx 0.15-0.2\%\,v_{\rm k}$). The MHD deviation described by the value $\beta_{\rm k}^{\rm MGD}$ varies over a wider range. The value of $\beta_{\rm k}^{\rm MGD}$ in general increases almost monotonically with the radial distance $r$ from $2\times 10^{-6}$ at the inner boundary of the disk to $10^{-2}$ ($\Delta v \approx 0.5\%\,v_{\rm k}$) at the outer boundary. The value of $\beta_{\rm k}^{\rm MGD}$ is significantly less than $\beta_{\rm k}^{\rm GD}$ in the region of $r\lesssim(3-10)$~au, that is, the contribution of magnetic tensions to centrifugal equilibrium is small compared to the contribution of the gas pressure gradient. This is due to the fact that the magnetic field is dynamically weak in this region, $\beta>1$, as Figure~\ref{fig:rad_mgd} shows. The magnetic field plays a dynamically important role, $\beta \sim 1$, in the outer part of the disk, and MHD deviation from the Keplerian rotation is of the order of the GD deviation, $\beta_{\rm k}^{\rm GD} / \beta_{\rm k}^{\rm MGD} \approx (1-2)$.

The simulation performed for the higher accretion rate, $\dot{M} = 10^{-7}\,M_\odot\,\mbox{yr}^{-1}$ (figure~\ref{fig:subkepler}b), shows that the magnetic field is dynamically strong both in the outer, and in the inner region of the disk in this case. The degree of the GD deviation ranges from $10^{-3}$ to $3\times 10^{-2}$ ($\Delta v\approx 1.5\%\,v_{\rm k}$), i.~e. it is slightly higher than in the case of $\dot{M}= 10^{-8}\,M_\odot\,\mbox{yr}^{-1}$. The MHD deviation, $\beta_{\rm k}^{\rm MGD}$, is comparable to the GD one both in the inner region of the disk, $r<0.2$~au, and in its outer part, $r>10$~au.

\subsection{Vertical structure of the disk}
\label{sec:vertical}
As follows from the system of equations of magnetostatic equilibrium (\ref{eq:motion_z3}-\ref{eq:induction_phi}), the influence of the magnetic field on the vertical structure of the disk is determined by the vertical gradient of magnetic pressure. \cite{cpmj21} have shown that equation (\ref{eq:induction_phi}) for the dominant azimuthal component of the magnetic field can be solved analytically. In the case of the Dirichlet boundary condition~(\ref{eq:BC_I}), when the magnetic field strength at the disk's surface is fixed,
 \begin{equation}
 B_{\varphi}(r,\,z) = B_{\rm ext}\frac{z}{z_{\rm s}} + \frac{1}{4}\frac{v_k z}{\eta}B_z\left[\left(\frac{z}{r}\right)^2 - \left(\frac{z_{\rm s}}{r}\right)^2\right].\label{eq:Bphi_I}
 \end{equation}
In this case, the vertical profile of $B_\varphi$ is, generally speaking, non-monotonic, so that the maximum value is achieved inside the disk.

In the case of the Neumann boundary condition~(\ref{eq:BC_II})
 \begin{equation}
B_{\varphi}(r,\,z)  = B_z \frac{v_{\rm k} z}{\eta}\left(\frac{z_{\rm s}}{r}\right)^2 \left[\frac{1}{4}\left(\frac{z}{z_{\rm s}}\right)^2 - \frac{3}{4}\right],\label{eq:Bphi_II}
\end{equation}
that is, $B_\varphi$ monotonically increases from zero in the equatorial plane to the maximum value at the disk's surface.

The choice of an appropriate boundary condition at the disk's surface is determined by the model of the medium above the disk. Both cases are analysed in this paper.

\subsubsection{Vertical structure at different radial distances}
Consider the vertical structure of the disk simulated with typical parameters at radial distances $r=0.2$, $1$ and $50$~au. Table~\ref{tab:z} shows the magnetic Reynolds number $R_{\rm m}$, the intensity of the vertical magnetic field component $B_z$, gas density and temperature in the equatorial plane at selected distances.  The magnetic Reynolds number is defined as $R_{\rm m}=v_{\rm k}H/\nu_{\rm m}$. The distance $r=0.2$~au corresponds to the thermal ionization region near the star, $r=1$~au lies inside the `dead' zone, $r=50$~au corresponds to the region outside the `dead' zone at the disk's periphery. As table~\ref{tab:z} shows, the magnetic field is frozen into the gas, $R_{\rm m}\gg 1$, outside the `dead zone', $r=0.2$ and $50$~au, while the diffusion of the magnetic field effectively develops, $R_{\rm m}\sim 1$, inside it.

\begin{figure*}
\includegraphics[]{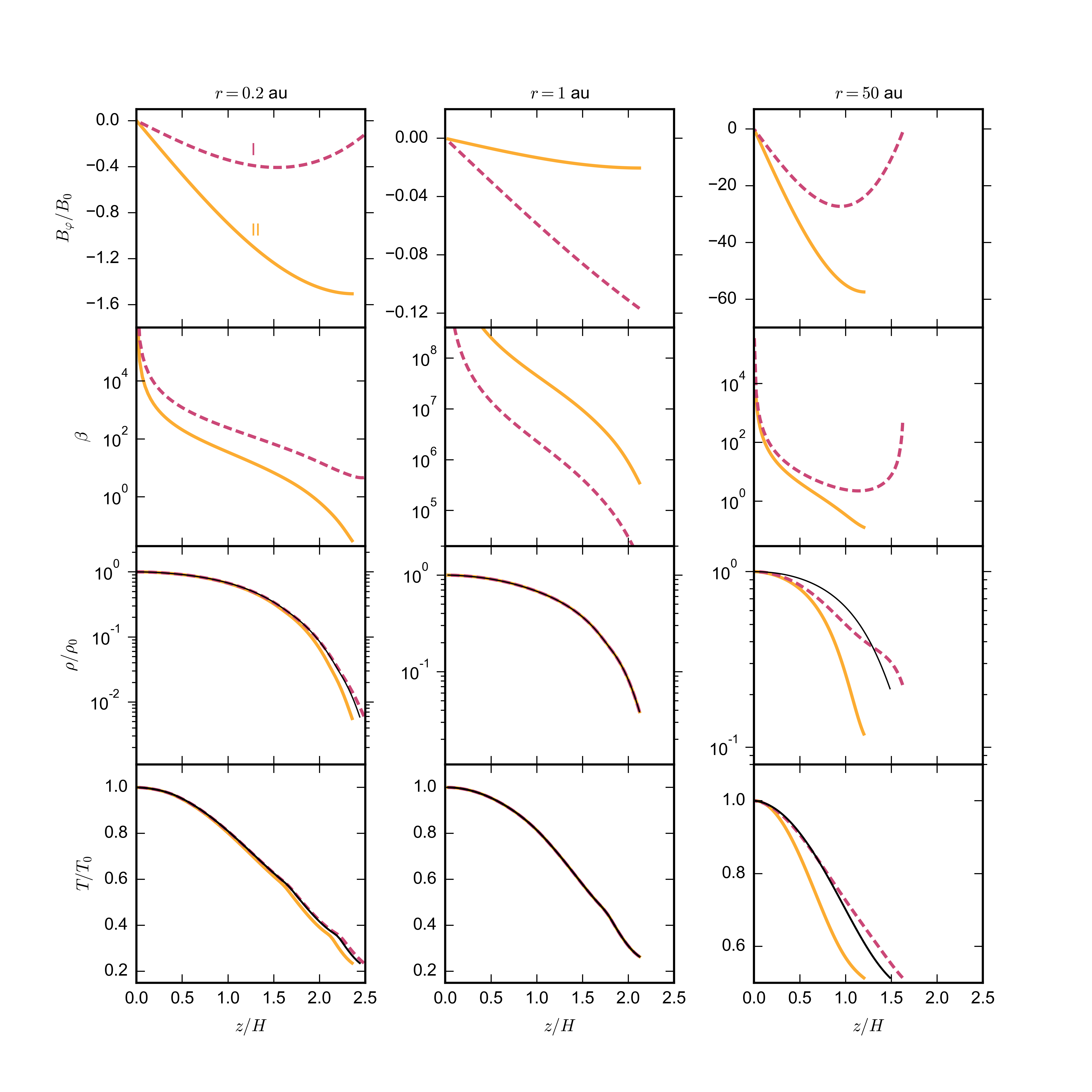}
\caption{The vertical structure of the disk at radial distances $r=0.2$~au (left column of panels), $r=1$~au (middle column), $r=50$~au (right column). The first row of panels: vertical profiles of the azimuthal magnetic field component  calculated for type I boundary conditions (dashed purple line marked as `I') and type II boundary conditions (solid yellow line marked as `II'). The second, third and fourth rows of panels: corresponding vertical profiles of the plasma beta, gas density and temperature. Thin black lines show vertical density and temperature profiles for the hydrostatic case ($B_\varphi=0$).}
\label{fig:z}
\end{figure*}

Figure~\ref{fig:z} shows vertical profiles of the azimuthal component of the magnetic field, plasma beta, gas density and temperature at selected distances. All values are shown in dimensionless form for the convenience of comparison. The intensity of $B_z$ (column 3 in table~\ref{tab:z}), gas density and temperature in the equatorial plane (columns 4 and 5 in table~\ref{tab:z}) are chosen as the corresponding unit scales.

Figure~\ref{fig:z} shows that the character of  $B_\varphi(z)$ profile depends on the boundary conditions at the surface and on the distance $r$.  
In the case of type I boundary conditions, profile $B_\varphi(z)$ is nonmonotonic outside the `dead' zone, $r=0.2$ and $50$~au.  For example, at $r=0.2$~au, the intensity of $B_\varphi$ varies from 0 in the equatorial plane to $0.1\,B_z\approx 0.7$~G at the surface, and takes the maximum value of $\approx 0.4\,B_z\approx 0.28$~G at altitude of $z\sim 1.5\,H$. In the outer part of the disk, $r=50$~au, the maximum value of $|B_\varphi|$ is $\approx 20\, B_z$, i.e.  the quasi-azimuthal magnetic field, $|B_\varphi|\gg B_z$ is most efficiently generated in this region. Magnetic field is practically not generated inside the `dead' zone due to effective Ohmic dissipation, and $B_\varphi$ increases almost linearly to the maximum value of $B_{\rm ext}=0.1\,B_z=5.5\times 10^{-3}$~G at the disk's surface.

In the case of type II boundary conditions, the intensity of $B_\varphi$ increases monotonically from the equatorial plane up to the surface of the disk. The generation of the magnetic field occurs most efficiently outside the `dead' zone, $r=0.2$ and $50$~au, so that the intensity of $B_\varphi$ exceeds the intensity of the initial field $B_z$.  Magnetic field is practically not generated inside the `dead' zone, $|B_\varphi| \lesssim 0.02\,B_z$.

The vertical profiles of the plasma beta depicted in the second row of panels in Figure~\ref{fig:z} show that the magnetic field is dynamically weak, $\beta\gg 1$, near the equatorial plane, $z\lesssim(0.5-1)\,H$, in all the cases considered. When moving away from the equatorial plane, the plasma beta decreases in all cases, except for the run with type II boundary conditions at $r=50$~au. In the latter case, the decrease of $\beta$ with height turns to an increase near the surface of the disk. The magnetic field is dynamically strong near the surface, $\beta\approx (0.1-10)$ at $z\gtrsim(1-1.5)\,H$ outside the `dead' zone, $r=0.2$ and $50$~au. The magnetic field is kinematic up to the surface of the disk inside the `dead' zone, $\beta\gg 1$.

\begin{table}
\centering
\caption[]{Parameters of the simulations of the vertical structure.\label{tab:z}}
\small
\begin{tabular}{ccccc}
\hline 
$r$, au & $R_{\rm m}$ & $B_z$, G & $\rho_0$, g~cm~$^{-3}$ & $T_0$, K \\ 
(1) & (2) & (3) & (4) & (5) \\ 
\hline 
$0.2$ & $194$ & $7.2$ & $2.79\times 10^{-9}$ & $1230$ \\ 
$1$ & $3$ & $5.5\times 10^{-2}$ & $1.43\times 10^{-9}$ & $350$ \\ 
$50$ & $5400$ & $4.9\times 10^{-3}$ & $2.13\times 10^{-13}$ & $13$ \\ 
\hline 
\end{tabular} 
\end{table}

Consider the influence of the magnetic field on the disk's structure. The vertical density profiles at selected distances (the third row of panels in Figure~\ref{fig:z}) show that the density distributions calculated taking into account the influence of the magnetic field differ from the hydrostatic ones outside the `dead' zone, $r=0.2$ and $50$~au. There are no differences inside the `dead' zone, $r=1$~au. The change in the $\rho(z)$ profiles is due to the effect of the magnetic pressure gradient in the regions where the magnetic field is dynamically strong. The sign of the magnetic pressure gradient depends on the boundary condition at the surface.

In the case of type I boundary conditions, the density profiles become flatter near the disk's surface and there is an increase in the characteristic half-thickness of the disk ($z$-coordinates of the photosphere) outside the `dead' zone, $r=0.2$ and $50$~au. As Figure~\ref{fig:z} shows, the increase in the half-thickness of the disk is small at $r=0.2$~au and substantial at $r=50$~au. In the latter case, not only the half-thickness of the disk increases, but also the characteristic slope of  $\rho(z)$ profile inside the disk changes. The increase in the half-thickness of the disk is due to the fact that the  magnetic field strength increases with height in the surface layers, so that the magnetic pressure gradient acts against the $z$-components of the gravitational force of the star and leads to the `expansion' of the disk in the regions with $\beta\sim 1$.

In the case of type II boundary conditions, the density profile near the disk surface is steeper than the hydrostatic one, and the corresponding half-thickness of the disk is smaller. This effect is most pronounced at $r=50$~au, where $z_{\rm s}\approx 1.5\,H$ in the hydrostatic case, and $z_{\rm s}\approx 1.2\,H$ in the magnetostatic case. The decrease in the characteristic half-thickness of the disk is due to the fact that the magnetic field strength monotonically increases towards the surface of the disk, so that the $z$-component of the electromagnetic force is co-directed with the $z$-component of the gravitational force of the star and leads to the `compression' of the disk in the regions with $\beta\sim 1$.

\subsubsection{Influence of the magnetic field on the thickness on the disk}

The previous section shows that magnetic pressure can change the thickness of the AD. Let us consider this effect in the application to the accretion disk of the solar mass T Tauri star with typical parameters corresponding to figure~\ref{fig:rad_fiduc}.

\begin{figure*}
\includegraphics[]{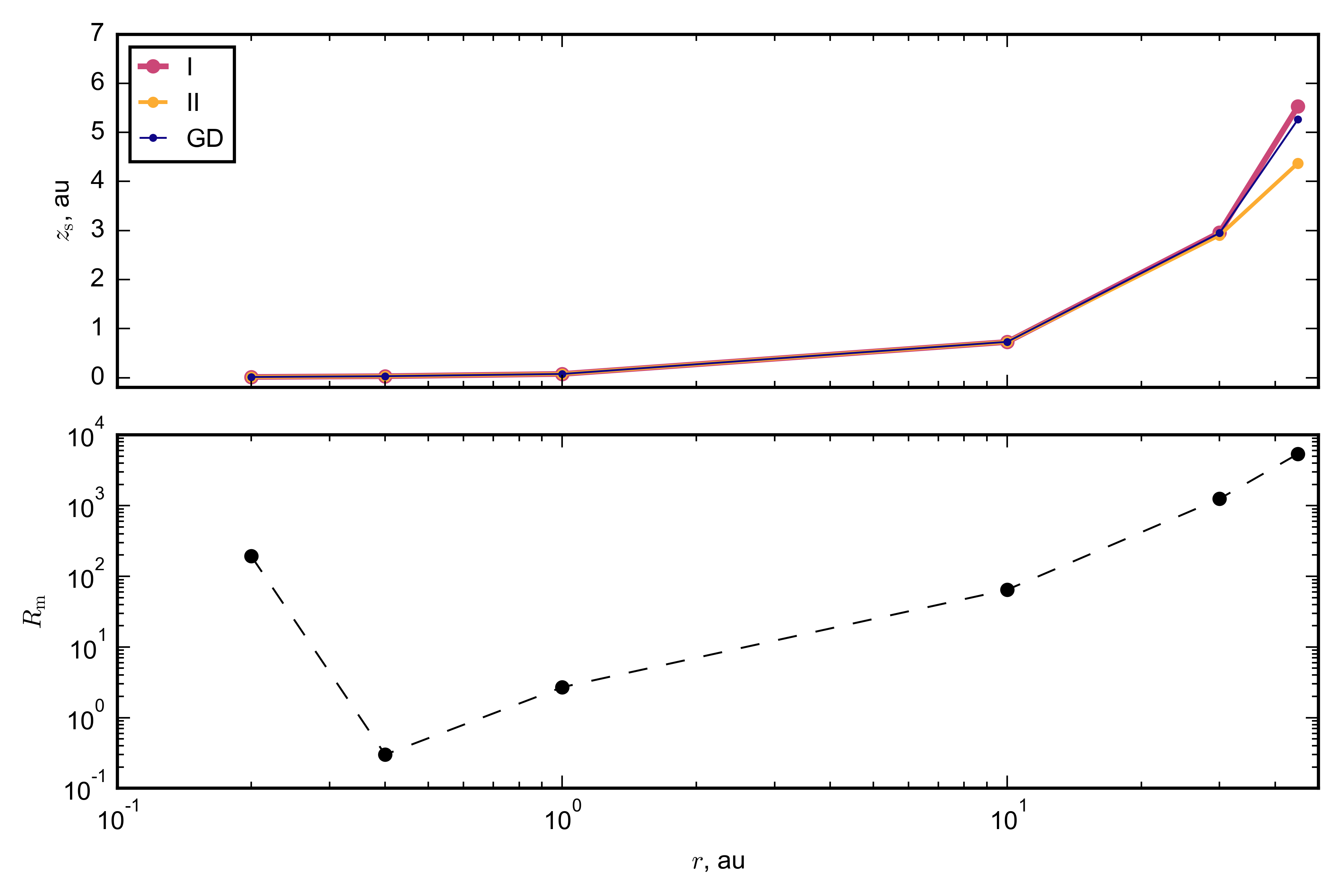}
\caption{Panel (a): dependence of the vertical coordinate of the photosphere of the disk, $z_{\rm s}$, on the radial distance $r$. The pink line with markers shows the case of type I boundary conditions for $B_\varphi(z)$, yellow line with markers corresponds to type II boundary conditions, blue line represents the simulation without the magnetic field. The simulations are carried out taking into account Ohmic dissipation. Panel (b): corresponding profile of the magnetic Reynolds number. The case of standard parameters is considered, corresponding radial structure of the disk is shown in Figure~\ref{fig:rad_fiduc}.}
\label{fig:zphot}
\end{figure*}

The upper panel of Figure~\ref{fig:zphot} shows the dependence of the coordinate of the disk's photosphere on the radial distance, determined on the basis of simulations of the vertical structure at $r=0.2$, $0.4$, $1$, $10$, $30$ and $50$~au. Three simulations are performed for each $r$: without taking into account the magnetic field (run `GD'), taking into account the magnetic field and Ohmic dissipation for boundary conditions of type I and II (runs `OD, I' and `OD, II' respectively). The values of the magnetic Reynolds number at each~$r$ are shown in the bottom panel of the figure.

Figure~\ref{fig:zphot} shows that the coordinate of the photosphere increases with $r$ from $z_{\rm s}\approx 0.015$~au at $r=0.2$~au to $z_{\rm s}\approx 5.5$~au at $r=50$~au in the simulation without magnetic field. Thus, the thickness of the disk increases with distance, but it remains geometrically thin, $z_{\rm s}\ll r$.

In the simulations with magnetic field, differences in profiles $z_{\rm s}(r)$ in comparison with run `GD' appear only in the outer region of the disk, $r>30$~au, where the magnetic Reynolds number is $R_{\rm m}\gtrsim 10^3$. In the case of type I boundary conditions, the magnetic pressure gradient leads to the thickening of the disk, which is manifested in the fact that the photosphere in this region of the disk is higher than in the `GD' case. In the case of type II boundary conditions, on the contrary, the gradient of the magnetic field leads to compression of the disk, and the photosphere is located lower than in run `GD'. The differences in the thickness of the disk with magnetic field from the hydrostatic one are indistinguishable in the inner region of the disk, $r\sim 0.2$~au, where $R_{\rm}\sim 200$.

Profiles $z_{\rm s}(r)$ in runs `GD' and `OD' coincide, i.e. the magnetic field does not affect the vertical structure of the disk in the region $r\in [0.3,\,30]$~au, corresponding to the `dead' zone (see section~\ref{sec:radial} and figure~\ref{fig:rad_fiduc}). Inside the `dead' zone, the magnetic Reynolds number $R_{\rm}$ is small, i.~e. effective Ohmic dissipation prevents the generation of a dynamically strong magnetic field.

In the considered cases, the maximum degree of thickening of the disk in the case of type I boundary conditions and compression in the case of type II boundary conditions is approximately 5 and 20~\%, respectively.

\section{Conclusions and discussion}
\label{sect:end}

In this paper, we develop our MHD model of the ADs of young stars to account for the dynamic effect of the fossil large-scale magnetic field on the disk's structure. The equations describing the radial structure of the disk are written taking into account the influence of magnetic tensions on the rotation of the gas. The vertical structure of the disk is determined from the magnetostatic equilibrium equation taking into account the magnetic pressure gradient. It is assumed that the surface of the disk lies in the region of its photosphere.

With the help of the developed model, we carried out simulations of the structure of the accretion disk of a solar mass T Taury star for different values of the accretion rate $\dot{M}$ and the radius of the dust grains $a_{\rm d}$. The simulations of the radial structure were performed under the assumption that the magnetic field strength in the disk is limited by the value corresponding to the equipartition between the thermal energy of the gas and magnetic energy.  The vertical structure of the disk was simulated for both type I and type II boundary conditions.

The simulations of the radial structure of the disk show that the magnetic field in the disk is kinematic for typical parameters, $\dot{M}=10^{-8}\,M_\odot\,\mbox{yr}^{-1}$ and $a_{\rm d}=0.1\,\mu$m: the plasma beta $\gg 1$, and the electromagnetic force does not affect the centrifugal equilibrium of the gas in the disk. Note that the plasma beta varies widely from $\sim 100$ near the inner edge of the disk, to $\sim 10^3-10^5$ inside the `dead' zone and $\sim 10$ at the periphery of the disk.

There is no `dead' zone, i.~e. the magnetic field is frozen into the gas throughout the disk in the case of the standard accretion rate, $\dot{M}=10^{-8}\,M_\odot\,\mbox{yr}^{-1}$, and large dust grains, $a_{\rm d}\gtrsim 1$~mm. In this case, the dynamically strong magnetic field with $\beta\sim 1$ is generated in the outer region of the disk, $r\gtrsim 30$~au. The tension of the fossil large-scale magnetic field in this region lead to decrease in the gas rotation velocity as compared to the Keplerian one. The degree of the deviation from the Keplerian velocity is $\sim 10^{-3}-10^{-2}$, which corresponds to $\Delta v = v_{\rm k} - v_\varphi\approx 0.15-0.5\%\,v_{\rm k}$. This value is comparable to the value due to the gas pressure gradient. The magnetic field is dynamically strong not only in the outer, $r\gtrsim 10$~au, but also in the inner region of the disk, $r\lesssim 0.2$~au at a higher accretion rate, $\dot{M}=10^{-7}\,M_\odot\,\mbox{yr}^{-1}$. The degree of the MHD deviation from the Keplerian rotation in this case is up to $\sim 3\times 10^{-2}$ ($\Delta v\approx 1.5\%\,v_{\rm k}$), which exceeds the degree of the gas-dynamic deviation for the selected parameters. Modern observations of several protoplanetary disks indicate the sub-Keplerian rotation in the outer regions of the studied disks with maximum deviation of $\Delta v$ up to several percent of $v_{\rm k}$~\citep{pinte18, dullemond20, teague21}.  The results obtained in our work show that this observed effect can be caused not only by the gradient of gas pressure, but also by magnetic tensions.

The deviation of the gas azimuthal velocity from the Kepler velocity is the main cause of the radial drift of dust grains and small bodies in PPDs~\citep{w77}. According to our results, the radial drift of dust particles in regions with dynamically strong magnetic field will occur at a higher speed than predicted by gas-dynamic simulations . This effect may be important from the point of view of the conditions of formation and dynamics of planetesimals in PPDs. The problem of the radial drift, taking into account the influence of the fossil magnetic field, requires further detailed research.

Simulations of the vertical structure of the accretion disk with standard parameters show that the behaviour of the profiles of the azimuthal component of the magnetic field depends on the type of boundary conditions and on the radial distance $r$.

In the case of type I boundary conditions corresponding to fixed intensity $B_{\rm ext}$ of the magnetic field at the disk's surface,  $B_\varphi$  profile is nonmonotonic outside the `dead' zone, and maximum  magnetic field strength is reached inside the disk. The most intense magnetic field generation occurs in the part of the disk external to the `dead' zone, where $|B_\varphi|\sim 20\,B_z$. Simulations show that maximum  magnetic field strength does not depend much on $B_{\rm ext}$ in this region. There is practically no magnetic field generation inside the `dead' zone, and $B_\varphi$ grows almost linearly with height remaining negligible compared to~$B_z$.

In the case of type II boundary conditions,
the magnetic field strength monotonically increases from the equatorial plane and reaches its maximum value at the surface of the disk. In the inner region, $r<1$~au, $|B_\varphi|\sim B_z$, that is, the magnetic field is quasi-azimuthal. The magnetic field is also quasi-azimuthal, with $|B_\varphi|\gg B_z$ in the outer region of the disk, $r\gtrsim 30$~au. The magnetic field generation does not occur even near the surface of the disk inside the `dead' zone, $|B_\varphi| \ll B_z$.

In both cases, the magnetic field is kinematic, $\beta\gg 1$, near the equatorial plane, $z\lesssim(1-1.5)\,H$, and dynamical, $\beta=0.1-10$, in the surface layers of the disk outside the `dead' zone. In the case of boundary conditions of the first kind, the magnetic pressure gradient leads to the increase in the characteristic thickness of the disk outside the `dead' zone. In the case of type II boundary conditions, magnetic pressure leads to decrease in the thickness of the disk, i.~e. to its `compression'.  Generally speaking, both cases are possible depending on the conditions above the disk. With typical parameters, the deviation of the disk thickness from the hydrostatic one is 5--20\,\% in the outer region, $r>30$~au.
For a detailed study of this effect, the model of the disk needs to be supplemented by the the model of the medium above the disk.

It should be noted that the generation of intense toroidal magnetic field in the surface layers of the accretion disk should inevitably lead to the development of magnetic buoyancy instability, also known as Parker instability~\citep[see][]{ParkerBook} or the Rayleigh-Taylor magnetic instability. For the development of the interchange mode of instability, it is sufficient that the magnetic field decreases with height more slowly than the density~\citep{tser60, hughes85}. It is also possible development of the undular mode if the magnetic field strength decreases with height. As our simulations show, both cases are realized in the surface layers of the disk outside the `dead' zone. A detailed study of the magnetic buoyancy instability in the accretion disks of young stars is planned in one of our next papers.

Further we plan to develop a fully self-consistent two-dimensional MHD disk model that takes into account the transfer of angular momentum by the large-scale magnetic field, as well as the inhomogeneity of the diffusivity and the magnetic buoyancy in the magnetostatics equations. The development of such a model will allow to take into account the effects associated with the formation of outflows from the disk and to determine the relative role of various angular momentum transport mechanisms in the ADs and PPDs of young stars.

{\bf Acknowledgments.}
The authors thank anonymous referee for some useful comments. The work of S.~A. Khaibrakhmanov in section~\ref{sec:radial} was carried out with the support of the Government of the Russian Federation and the Ministry of Higher Education and Science of the Russian Federation under grant 075-15-2020-780 (N13.1902.21.0039, contract 780-10). The work of A.~E.~Dudorov in the section~\ref{sec:vertical} was supported by the Russian Foundation for Basic Research (project No. 20-42-740013). S.~A. Khaibrakhmanov  thanks S.~N.~Zamozdra for some useful comments.

\bibliographystyle{mnras}
\bibliography{pik} 

\end{document}